\documentstyle[prd,aps,epsf,epsfig]{revtex} 
\begin{document}
\draft
\twocolumn[\hsize\textwidth\columnwidth\hsize\csname@twocolumnfalse\endcsname
%
\title{Improved percolation thresholds for rods in three-dimensional boxes}
%
\author{R. Florian$^a$ and Z. N\'eda$^b$}
\address{ $^a$Center for Cognitive and Neural Studies,
          str. Saturn 24, RO-3400 Cluj-Napoca, Romania \\
          \texttt {florian@coneural.org}  \\
          $^b$Babe\c{s}-Bolyai University, Dept. of Physics, 
          str. Kog\u{a}lniceanu 1, RO-3400 Cluj-Napoca, Romania \\
          \texttt {zneda@phys.ubbcluj.ro}}

\maketitle

\begin{abstract}

We improve our previous results for the percolation thresholds of isotropically 
oriented rods in three dimensional boxes. We prove again the applicability of 
the excluded volume rule in the slender-rod limit (radius/length$\rightarrow0$).
Other limits for the rod sizes are discussed and 
important finite-size effects are revealed.

\end{abstract}

\pacs{PACS numbers: 64.60.Ak, 02.50.Ng, 05.40.+j  }
\vspace{2pc}
]

\vspace{1cm}

We have previously \cite{Eprint} revealed errors of earlier simulations \cite{BalBin}
for the percolation of permeable rods in three dimensions. We have also presented
corrected results \cite{PhysRev} that confirm the excluded volume theory. In this 
short preprint we give more accurate numerical values for the percolation  thresholds, 
obtained by extended computer simulations. Given the importance of such results in
designing composite materials (electrically conducting polymers, fiber-inforced plastics)
it is useful to summarize and disscus the obtained percolation thresholds.  

The studied system is formed by permeable sticks with the form of capped cylinders (cylinders 
of length $L$ and radius $R$ capped with two hemispheres of radius $R$).
It was conjectured \cite{Bb3} that 
the percolation threshold, $q_p$, is proportional 
to the inverse of the expected excluded volume, $V_{ex}$: 
\begin{equation}
q_p=\frac{N_c}{V}\propto\frac{1}{V_{ex}} 
\end{equation}
(We denoted by $N_c$ the number of sticks at percolation and by $V$ the 
volume of the cube in which the percolation problem is considered.)
For sticks with the form of capped cylinders the excluded volume is given by
\begin{equation}
V_{ex}=\frac{32 \pi}{3} R^3 + 8 \pi L R^2 + 4 L^2 R <sin(\gamma)>,
\end{equation}
where $<sin(\gamma)>$ is the average value of $sin(\gamma)$ for two
randomly positioned sticks, and $\gamma$ is the angle between them.
For isotropic orientation of rods ($<sin(\gamma)>=\pi/4$) it was
shown by a cluster expansion method \cite{Bug} that the proportionality
in (1) becomes equality in the $R/L\rightarrow 0$ slender-rod limit.
This equality is known as the excluded-volume rule.

We have studied the problem inside a cube with unit sizes.
In order to preserve homogeneity near the cube's frontiers, the coordinates of
the centers of the cylinders were generated uniformly in the interval
$[-(L/2+R),1+(L/2+R)]$. The orientation of the cylinders was
isotropic. We detected the intersection of two capped cylinders by 
calculating the minimum distance between points on the two axes of the 
corresponding cylinders and by comparing this distance with $2R$. 
Each time a new stick was generated, it was assigned to a cluster 
if it intersected a previous one, otherwise a new cluster was created; a stick may also unify two or more clusters
that were previously separated. We considered  
percolation achived when a cluster spanned the cube from one face 
to the opposite one. The critical concentration is given by 
the number 
of sticks inside the cube at percolation, $N_c$. 
If a capped cylinder was only
partially inside the cube, it contributed to $N_c$ with a fractional value less 
than one, corresponding to the fraction of its volume inside the cube to its 
total volume. We  generated many percolation realizations (usually $5000$) for each considered 
$L$ and $R$ pair, and 
the average $N_c$ was calculated as the mean of the obtained percolation thresholds.

The obtained results are summarized in Figs. 1 - 3.

\begin{figure}[h]
\epsfig{figure=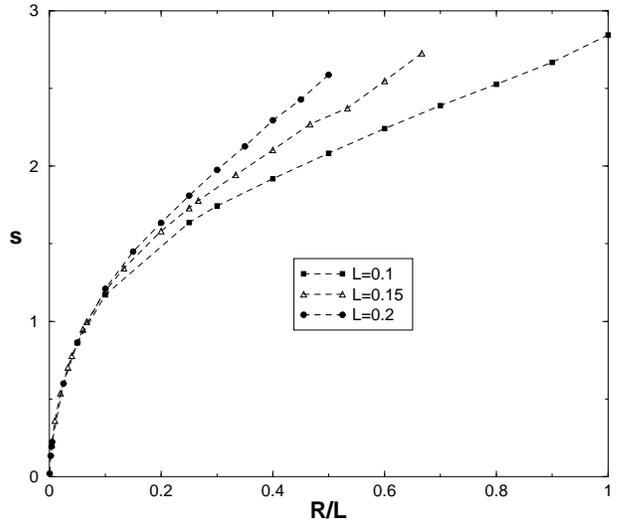,width=2.8in,angle=-90}
\caption{$s=q_pV_{ex}-1$ as a function of the $R/L$ aspect ratios
of the sticks. Data  for three different stick lengths, $L$, are presented.}
\label{fig1}
\end{figure}

On Fig. 1 we plot
the quantity $s=q_pV_{ex}-1$ as a function of $R/L$ for various fixed
$L$ values. In the limit $R/L\rightarrow 0$ our simulations suggest
the analytically predicted $s=0$ relation \cite{Bug}.
This is nicely proven by our large-scale simulation
data for $L=0.2$ (for systems of up to $N_c=8 \cdot 10^4$ rods). 
The convergence for the applicability of the excluded volume rule 
is however rather poor. As an example in this sense we
found in the $R/L\rightarrow 0$ limit ($R/L<0.05$), for $L=0.15$, $s$ scaling
as a function of $R/L$ with an exponent of $0.58$.  
From Fig. 1 is
also clear that for smaller values of $L$ and the same $R/L$ ratios the
value of $s$ gets smaller. There are thus important finite-size
effects,  which are less evident in the $R/L\rightarrow 0$ limit.

We have checked that in the limit of $L\rightarrow 0$ the $s=0$ equality
still does not hold. This is clear from our simulation
data for $R/L=0.5$, $R/L=0.25$ and $R/L=0.1$. The data presented on Fig. 2 suggest
that in the limit $L\rightarrow 0$, $s$ is linearly converging to 
$1.58$, $1.46$ and $1.14$ for $R/L=0.5$, $R/L=0.25$ and $R/L=0.1$, respectively.
This linear convergence may be useful for approximating the $s$ value given $R$ and $L$.

\begin{figure}[h]
\epsfig{figure=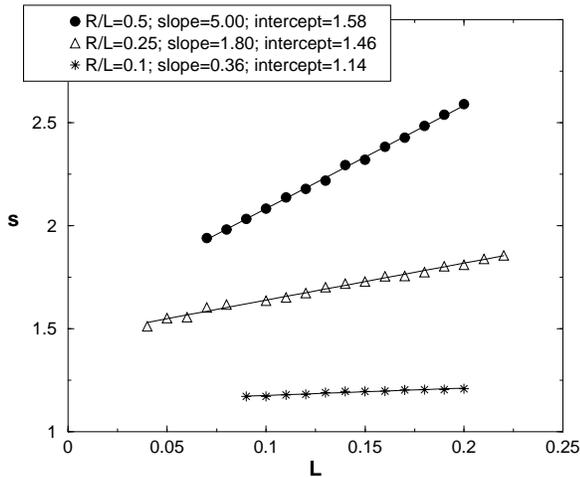,width=3.0in,angle=0}
\caption{ Finite size effects: $s=f(L)$  for two different
aspect ratios of the sticks. Continuous lines are the best linear
fit. In the $L\rightarrow 0$ limit we got $s=1.58$, $s=1.46$ and $s=14$ 
for $R/L=0.5$, $R/L=0.25$ and $R/L=0.1$, respectively. }
\label{fig2}
\end{figure}

Problems with the applicability of the excluded-volume rule 
appears also in the $L\rightarrow 1$ limit, where the length of
the sticks approaches the charactheristic length of the box. This is nicely visible
from Fig.~3 where we have plotted $s$ as a function of $R/L$ for two fixed
$R$ values ($R=0.01$ and $=0.02$, respectively). In the limit of very small $R/L$ values
one can observe again the increasing difference from the $s=0$ theoretical result, due to $L$
values approaching the characteristic box size.

\begin{figure}[h]
\epsfig{figure=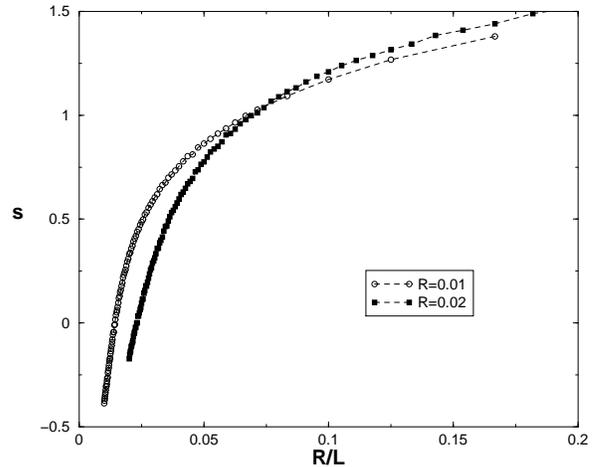,width=2.5in,angle=-90}
\caption{ $s=f(R/L)$ for two different
radii of the sticks. }
\label{fig3}
\end{figure}

In conclusion, our results suggest that the excluded-volume rule is applicable
in the slender-rod limit ($R/L\rightarrow 0$) for stick lengths much smaller 
than the characteristic box-size. For all other cases important deviations
from this rule are present. Decreasing the rod length improves the 
applicability of the excluded-volume rule, but the $R/L \rightarrow 0$ condition is 
still necessary in the $L \rightarrow 0$ limit.


\begin{references}
\bibitem{Eprint} Z. Neda, R. Florian; cond-mat/9804326 (1998)
\bibitem{BalBin} I. Balberg, N. Binenbaum and N. Wagner; Phys. Rev. Lett. {\bf 52}, 
1465 (1984)
\bibitem{PhysRev} Z. Neda, R. Florian and Y. Brechet; Phys. Rev. E {\bf 59}, 3717 (1999)
\bibitem{Bb3} I. Balberg, C.H. Anderson, S. Alexander and N. Wagner; Phys. Rev. B
{\bf 30}, 3933 (1984)
\bibitem{Bug} A.L.R. Bug, S.A. Safran and I. Webman; Phys. Rev. Lett. {\bf 54}, 
1412 (1985)
\end{references}
\end{document}